\documentclass[prl,aps,twocolumn]{revtex4}
\usepackage{graphicx}
\usepackage{dcolumn}
\usepackage{bm}

\begin{document}


{\bf Comment on "Tunnel Window's Imprint on Dipolar Field
Distributions"}

\vspace{0.5cm}

Some years ago it was predicted that the tunneling relaxation of
an ensemble of tunneling nanomagnets or magnetic molecules would
be governed simultaneously by their interactions with nuclear
spins \cite{PS96} and their mutual dipolar interactions
\cite{PS98}. Experimentalists have interpreted measurements in
various tunneling magnetic systems using this theory, eg., in
$Fe$-8 molecules- where evidence for the role of the nuclear spins
was found \cite{ww00}. In a recent letter Alonso et al.
\cite{FernHole} present Monte Carlo simulations for interacting
dipoles. Short-time relaxation was caused by a simple noise field,
acting uniformly over an "energy window" of width $2\delta h_{hf}$
(to simulate the fluctuating nuclear spin bias). A "hole"
developed in the magnetisation distribution $M(\xi)$ over the
longitudinal bias field $\xi$, of half-width $W \approx 0.75 \;
\delta h_{hf}$. Alonso et al. thereby claim to establish a
connection between the experimental short-time holewidth $2 \xi_o$
and $2\delta h_{hf}$.

However the original predictions of the Prokof'ev Stamp (PS)
theory, including the hole \cite{PS98}, relied on different
physics. The nuclear dynamics is driven by 2 processes, viz (i)
intrinsic nuclear bath fluctuations, and (ii) that dynamics of the
nuclear spins which is {\it stimulated} by the molecular spin
flips. Process (i) does cause longitudinal noise, but {\it not}
across the whole multiplet of nuclear spin states. {\it However},
molecular flips can occur even off resonance (process (ii)), if
transitions are simultaneously stimulated in the nuclear bath,
exchanging energy with the tunneling molecule. The importance of
this latter process is measured in the PS theory \cite{PS96,PS00}
by a parameter $\kappa$. If $\kappa \sim O(N^{1/2})$ or greater,
where $N$ is the number of nuclear spins in the molecule, the
theory predicts an initial hole half-width $\xi_o \sim E_o$, where
$E_o$ is the half-width of the nuclear spin multiplet (see ref.
\cite{PS00}, pp 698-701, and 704). This half-width $E_o$ is {\it
not} a "typical hyperfine coupling", as asserted in
\cite{FernHole}, but much bigger; in fact $E_o^2 = \sum_k
(\omega_k^{\parallel})^2 (I_k+1)I_k/3$, where $I_k$ is the $k$-th
nuclear spin, and in low fields $\omega_k^{\parallel}$ is the
hyperfine coupling to the $k$-th nuclear spin. At longer times,
the hole width and shape is governed in the PS theory not by
nuclear spins, but by the time evolution of dipolar fields
\cite{PS98}.

Fig. 1 (at bottom) shows our computation of $\kappa({\bf
H}_o^{\perp})$ for a $Fe$-8 molecule in low transverse field ${\bf
H}_o^{\perp}$, where the dipolar hyperfine couplings are easily
calculated. Since experiments have average intermolecular fields
$\sim 200-300~ Oe$, we see that in the experiments $\kappa$ is
large, and thus one expects $\xi_o$ to be not much less than $E_o$
(as found experimentally- see top left inset in Fig 1). We also
show the holewidth calculated using Monte Carlo simulations for a
sample which is strongly annealed, but here using the kinetic
equation of ref. \cite{PS98}, rather than just a phenomenological
noise.

The 2 approaches give different results for the hole evolution.
The PS theory unambiguously predicts that $\xi_o \sim E_o$, but
only for large $\kappa$- in agreement with the experiments, using
different nuclear isotopes, done on $Fe$-8. A further check will
come in transverse fields $>2-3T$, where $\kappa$ falls steeply,
and so we predict a steep fall of $\xi_o$ (which would be very
hard to understand in the approach of ref. \cite{FernHole}).

\vspace{0.7cm}
\begin{figure}[h]
\centering 
\vspace{-2.3cm}
\hspace{0.cm}
\includegraphics[scale=0.4]{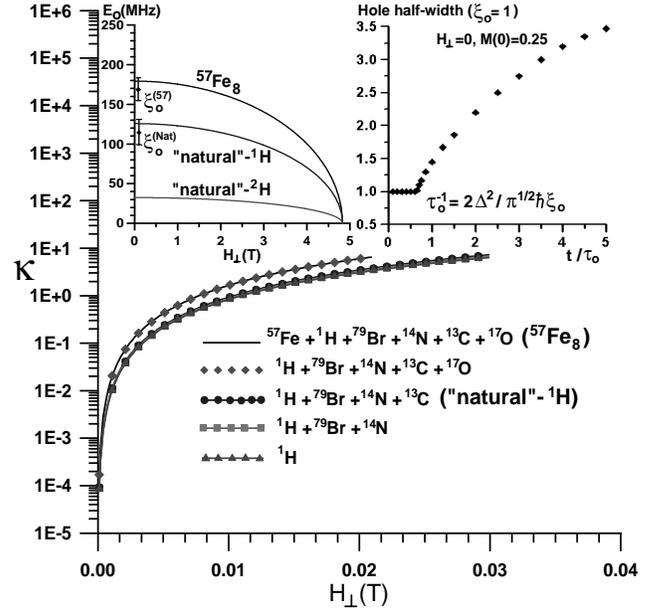}
\vspace{-1.5cm}
\caption{At bottom, the variation of $\kappa({\bf H}_o^{\perp})$ 
with transverse field ${\bf H}_o^{\perp}$ in $Fe$-8 
at low fields, for various isotopic mixtures. Top left inset shows 
calculated values of $E_o$ for 3 isotopic mixtures, and also 
experimental measurements of $\xi_o$ for two of these, from ref. 
\cite{ww00}. Top right shows the hole half-width (in units of 
$\xi_o$) for a sample with time (in units of $\tau_o$), using MC 
simulations on a cube of volume $40^3$ sites, and assuming large 
$\kappa$; for short times it is constant and equal to $\xi_o$, 
with $\xi_o = E_o$ for large $\kappa$.} 
\label{fig:fig1}
\end{figure}


\noindent
\begin{tabular}{l}
P. C. E. Stamp$^{1}$, and I. S. Tupitsyn$^{1,2}$ \\
$^{1}$ Physics Department, University of British Columbia, \\
6224 Agricultural Rd., Vancouver BC, Canada, V6T 1Z1. \\
$^{2}$ Russian Research Center "Kurchatov Institute", \\
Moscow 123182, Russia. $\;\;\;$ \\
\end{tabular}

\end{document}